\documentstyle[12pt,axodraw,epsfig]{article}

\def\beq{\begin{equation}}
\def\eeq#1{\label{#1}\end{equation}}
\def\eeqn{\end{equation}}
\def\beqa{\begin{eqnarray}}
\def\eeqa#1{\label{#1}\end{eqnarray}}
\def\eeqan{\end{eqnarray}}
\def\CR{\nonumber \\ }
\def\leqn#1{(\ref{#1})}

\def\mm{\tilde{m}_1}
\def\mp{\tilde{m}_2}

\def\t{\theta_t}
\def\tl{\tilde{t}_1}
\def\th{\tilde{t}_2}

\def\Mstop{M_{\tilde{t}}}
\def\MS{M_{\rm susy}}

\newcommand{\bspace}{\!\!\!\!}
\def\met{\mbox{$E{\bspace}/_{T}$}}

\def\stacksymbols #1#2#3#4{\def\theguybelow{#2}
    \def\vp{\lower#3pt}
    \def\sp{\baselineskip0pt\lineskip#4pt}
    \mathrel{\mathpalette\intermediary#1}}

\def\intermediary#1#2{\vp\vbox{\sp
     \everycr={}\tabskip0pt
     \halign{$\mathsurround0pt#1\hfil##\hfil$\crcr#2\crcr
              \theguybelow\crcr}}}

\def\gsim{\stacksymbols{>}{\sim}{2.5}{.2}}
\def\lsim{\stacksymbols{<}{\sim}{2.5}{.2}}

\begin{document}

\begin{titlepage}
\begin{flushright}
{\tt hep-ph/yymmnnn} \\
\end{flushright}

\vskip.5cm
\begin{center}
{\huge \bf A Collider Signature of the } \\
\vskip0.4cm
{\huge \bf Supersymmetric Golden Region} 
\vskip.2cm
\end{center}
\vskip1cm

\begin{center}
{\bf Maxim Perelstein and Christian Spethmann} \\
\end{center}
\vskip 8pt

\begin{center}
{\it Cornell Institute for High-Energy Phenomenology, 
Cornell University, Ithaca, NY~14853}
\end{center}

\vglue 0.3truecm

\begin{abstract}
\vskip 3pt \noindent
Null results of experimental searches for the Higgs boson and the
superpartners imply a certain amount of fine-tuning in the
electroweak sector of the Minimal Supersymmetric Standard Model (MSSM).
The ``golden region'' in the MSSM parameter space is the region where the 
experimental constraints are satisfied and the amount of fine-tuning is 
minimized. In this region, the stop trilinear soft term $A_t$ is large,
leading to a significant mass splitting between the two stop mass 
eigenstates. As a result, the decay $\th\to\tl Z$ is kinematically
allowed throughout the golden region. We propose that the experiments 
at the Large Hadron Collider (LHC) can search for this decay through an 
inclusive signature, $Z+2j_b+\met+X$. We evaluate the Standard Model 
backgrounds for this channel, and identify a set of cuts that would 
allow detection of the supersymmetric contribution at the LHC for 
the MSSM parameters typical of the golden region. We also discuss other
possible interpretations of a signal for new physics in the  $Z+2j_b+\met+X$
channel, and suggest further measurements that could be used to distinguish 
among these interpretations.  
\end{abstract}

\end{titlepage}

\section{Introduction}

It is widely believed that physics at the TeV scale is supersymmetric. 
The simplest realistic implementation of this idea, the minimal supersymmetric 
standard model (MSSM)~\cite{HK,Martin}, is the most popular extension of the 
standard model 
(SM). However, null results of experimental searches for the superpartners and,
especially, the Higgs boson, place non-trivial constraints on the parameters 
of the model. 
Furthermore, the requirement that the observed electroweak symmetry breaking 
(EWSB) occur without significant fine-tuning places an additional constraint. 
It is well known that there is a certain amount of tension between these two 
constraints~\cite{LEPpar}. Several authors have interpreted this tension as 
a motivation to extend the minimal model~\cite{extend}, or to question the
conventional ideas about naturalness~\cite{danger}. An 
alternative interpretation, which we will explore in this paper, is that data 
and naturalness point to a particular ''golden'' region within the parameter 
space of the minimal model, where the experimental bounds are satisfied and 
fine-tuning is close to the minimum value possible in the MSSM.  
This minimal value itself depends on the messenger scale of
supersymmetry breaking $\Lambda_{\rm mess}$, determined by dynamics 
outside of the MSSM, in addition to the MSSM parameters\footnote{For example, 
it was claimed in Ref.~\cite{KN} that in models with ``mirage mediation'' of 
SUSY breaking~\cite{MM} the scale $\Lambda_{\rm mess}$ can be as low as 1 TeV, 
resulting 
in fine-tuning of 20\% or better. See Ref.~\cite{PT} for a 
discussion of difficulties in realizing such a scenario, and 
Ref.~\cite{antiPT} for an alternative implementation.}. However, for 
{\it any} 
$\Lambda_{\rm mess}$, the points in the golden region require {\it less} 
fine-tuning
compared to the rest of the MSSM parameter space. Thus, independently of 
the model of SUSY breaking, nature seems to provide us with a hint about
what the MSSM parameters might be\footnote{An explicit model of supersymmetry 
breaking in a grand-unified framework which naturally generates SUSY breaking
parameters in the golden region was constructed in~\cite{DKK}.}. In this 
paper, we will discuss  
experiments at the Large Hadron Collider (LHC) which will be able to determine
whether this hint is correct. 

Both the Higgs mass bound and naturalness considerations probe the 
effective Higgs potential, which is primarily determined by the 
parameters of the Higgs and top sectors of the MSSM. (The quantum part
of the potential is dominated by the top/stop loops due to a large value of 
the top Yukawa coupling.) It is therefore these sectors that are most
directly constrained by data. We will focus on collider measurements
probing these sectors.\footnote{Our approach is more model-independent than
that of Kitano and Nomura in Ref.~\cite{KNcoll}, which also considered 
collider signatures of the MSSM with parameters in the golden region. 
However, the signatures studied in~\cite{KNcoll} mainly 
probe the features of the superpartner spectrum dictated by a specific
(mirage mediation) model of SUSY breaking~\cite{KN}, rather than the direct 
consequences of data and naturalness.}  

The golden region is characterized by relatively small 
values of the $\mu$ parameter and the stop soft masses $m_{Q^3}$, $m_{u^{3}}$ 
(both are required to minimize fine-tuning of the $Z$ mass), and a large stop
trilinear soft term $A_t$ (required to raise the Higgs mass above the LEP2 
lower bound). The spectrum
is then expected to contain light neutralinos and charginos with a
substantial higgsino content, as well as two light (sub-TeV) stop mass 
eigenstates, $\tl$ and $\th$, with a large (typically a few hundered GeV) 
mass splitting. A striking consequence of such a ``split stop'' spectrum is 
that the decay 
\beq
\th \to \tl + Z
\eeq{cool}
is kinematically allowed. Observing this decay at the LHC would 
provide clear evidence that the stop mass difference is larger than the 
$Z$ mass, and studying the $Z$ distributions would provide an approximate 
measurement of this quantity. In this paper, we will argue that 
the decay~\leqn{cool} should be observable at the LHC, with realistic 
integrated luminosity, for the MSSM parameters in the golden region.

The experimental signature of the decay~\leqn{cool} depends on the
decay pattern of the $\tl$. Since stops are almost always pair-produced at 
the LHC, it also depends on how the second $\th$ decays. The details of
both decay patterns depend on the superpartner spectrum. However, 
both $\tl$ and $\th$ decay products always contain a $b$ quark, produced 
either directly or through a top decay, as well as (under the usual
assumptions of conserved R parity and weakly interacting lightest 
supersymmetric particle) large missing 
transverse energy. We therefore propose an inclusive final state
\beq
Z + 2 j_b + \met + X,
\eeq{signature}
where $Z$ is assumed to be reconstructed from leptonic decays and 
$j_b$ denotes a $b$ jet, as a signature of the $\th\th^*$ 
production followed by the decay~\leqn{cool}. 

Throughout the golden region of the MSSM, both the $\th$ pair-production cross 
section and the branching fraction of the decay~\leqn{cool} are sizeable. 
Therefore, a null result of a search for a non-SM contribution in the 
channel~\leqn{signature} would provide a strong argument against this
scenario. Unfortunately, a positive identification of non-SM physics in 
this channel would {\it not} necessarily imply that the stops are split.
Indeed, in the MSSM, events in this channel may appear even if 
the decay~\leqn{cool} is kinematically forbidden, since $Z$ bosons may
also be produced in decays of neutralinos and charginos~\cite{HB}. For 
example, a cascade
\beq
\tilde{b} \to b \chi_2^0,~~~\chi_2^0 \to Z \chi_1^0,
\eeq{casc}
or a similar cascade with charginos replacing the neutralinos, 
gives the signature~\leqn{signature}. Distinguishing these interpretations
is difficult, and there is no single ``silver bullet'' observable that 
would remove this ambiguity. However, a variety of measurements can be 
used to shed light on this question (see Section 5), and combining all
available evidence may allow one to build a convincing case for (or against)
the interpretation of the signature~\leqn{signature} in terms of the 
decay~\leqn{cool}.

The paper is organized as follows. In Section 2, we review the fine-tuning 
and Higgs mass constraints in the MSSM, as well as other experimental 
results that determine the shape of the golden region. In Section 3, we define 
a benchmark point which is characteristic of the golden region and suitable
for studying its collider phenomenology. Section 4 is dedicated to a detailed 
analysis of the observability of the $Z+2j_b+\met$ signature, including a 
study of the SM backgrounds. In Section 5, we discuss the alternative 
interpretations of this signature within the MSSM, and outline the 
measurements that would need to be performed to discriminate between these
interpretations. Section 6 contains our conclusions, and outlines some 
possible directions for future work. 

\section{The Golden Region}

In this section, we will discuss the constraints on the MSSM parameters
imposed by current experimental data and naturalness, focusing on the Higgs 
and top sectors. Our goal is to understand the
qualitative features of the MSSM golden region, rather than to determine the 
precise location of its boundaries which are in any case fuzzy due to an 
inherent lack of precision surrounding the concept of fine-tuning. With
this motivation, we will make several approximations which greatly clarify
the picture. 

Phenomenological studies of the MSSM are complicated by the large number of
free parameters. Typically, studies are performed within simplified 
frameworks, which assume certain correlations among the parameters motivated 
by high-scale unification and/or by specific models of SUSY breaking.
However, the shape of the golden region is to a great extent independent of
such assumptions. The Higgs sector of the MSSM is strongly coupled to the top 
sector, but couplings to the rest of the MSSM are weaker. One may therefore 
begin by considering the Higgs and top sectors in isolation; 
that is, the gauge and non-top Yukawa couplings are set to zero. In this
approximation, physics is described in 
terms of the holomorphic Higgs mass $\mu$ and the six parameters appearing 
in the soft Lagrangian for the Higgs and top sectors: 
\beq
{\cal L} \,=\, -m_u^2 |H_u|^2 -m_d^2 |H_d|^2 - \left(
b H_u^T H_d + {\rm c.c.}\right)
-m_{Q^3}^2Q^{3\dagger}Q^3-m_{u^{3}}^2|u^3|^2-
\left(y_t A_t Q^{3\dagger}H_u u^3 + {\rm c.c.}\right), 
\eeq{Lsoft}
where $y_t$ is the MSSM top Yukawa coupling, $y_t=y_t^{\rm SM}/\sin\beta$.
Since the model has to reproduce the known EWSB scale, $v=174$ GeV, 
only six parameters are independent. We choose the physical basis:
\beq
\tan\beta, \mu, m_A, \mm, \mp, \theta_t,
\eeq{parameters}
where $m_A$ is the CP-odd Higgs mass, $\mm$ and $\mp$ are stop eigenmasses
(by convention, $\mp>\mm$) and $\theta_t$ is the stop mixing 
angle. We will analyze the fine-tuning and Higgs mass constraints in 
this approximation and map out the golden region in the six-parameter 
space~\leqn{parameters}.

Before proceeding, let us discuss the sizes of contributions to
the relevant observables that are omitted in this approximation scheme.
The leading contributions to the Higgs effective potential due to the 
$SU(3)$, $SU(2)$ and $U(1)_Y$ gauge interactions and the bottom Yukawa 
coupling\footnote{The 
corrections due to other Yukawa couplings are always negligible.}
are expected to be of the order 
\beq
\frac{g_3^2 M_3^2}{16\pi^2 \lambda_t^2 \Mstop^2},~~~ 
\frac{g_2^2 M_2^2}{\lambda_t^2 \Mstop^2},~~~
\frac{g_1^2 M_1^2}{\lambda_t^2 \Mstop^2},~~~
\frac{m_b^2 M_{\tilde{b}}^2 \tan^2\beta}{m_t^2 \Mstop^2},
\eeq{corr}
respectively, compared to the one-loop top sector contribution. 
Here $g_i$ and $M_i$ are the gauge couplings and (weak-scale) gaugino masses 
for each group, and $\Mstop$ is the stop mass scale, 
which can be conveniently taken as the average between the two stop
eigenmasses. (The same definition can be made for $M_{\tilde{b}}$ if 
sbottoms are non-degenerate.) For a wide range of sensible 
superpartner spectra, these corrections are subdominant: this is the case if
\beq
M_1/\Mstop\lsim 4,~~M_2/\Mstop \lsim 2,~~ 
M_3/\Mstop \lsim 10,~~
M_{\tilde{b}}\lsim \frac{35 \Mstop}{\tan\beta}. 
\eeq{bottoms_up}
The following discussion is valid for spectra obeying these constraints.
If some of the above inequalities are violated, the analysis could be 
easily extended to include the corresponding effects; however, little 
additional insight would be gained.

\subsection{Constraints on the Higgs Sector}

\begin{figure}[tb]
\begin{center}
\includegraphics[width=6cm]{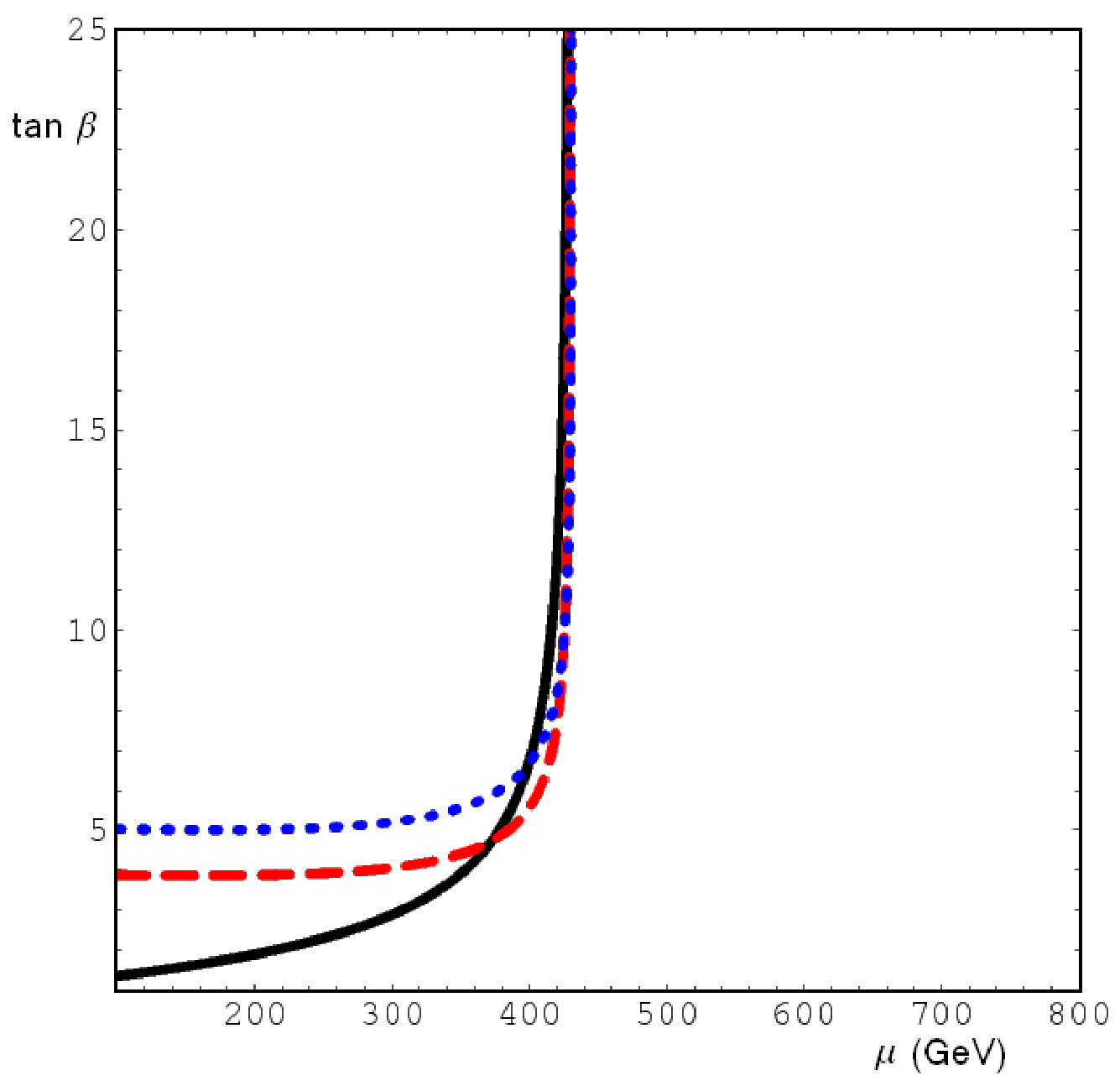}
\vskip2mm
\caption{Contours of 1\% fine-tuning in the ($\mu, \tan\beta$) plane.
The black (solid) contour corresponds to $m_A=100$ GeV, but remains 
essentially unchanged for any value of $m_A$ in the range between 100 and 
1000 GeV. The red (dashed) and blue (dotted) contours correspond to $m_A=1.5$ 
and 2 TeV, respectively.}
\label{fig:FT}
\end{center}
\end{figure}

At tree level, the $Z$ mass in the MSSM is given by
\beq
m_Z^2 \,=\, - m_u^2 \left( 1-\frac{1}{\cos 2 \beta} \right) - 
m_d^2 \left( 1+\frac{1}{\cos 2 \beta} \right)-2|\mu|^2 \,,
\eeq{zmass}
where
\beq
\sin 2 \beta  = \frac{2 b}{m_u^2 + m_d^2+2|\mu|^2}\,.
\eeq{sin}
Following Barbieri and Guidice~\cite{BG}, we quantify fine-tuning by 
computing
\beq
A(\xi)\,=\,\left| \frac{\partial\log m_Z^2}{\partial\log \xi}\right|,
\eeq{ftpars}
where $\xi=m_u^2, m_d^2, b, \mu$ are the relevant Lagrangian parameters.
In terms of the physical parameters~\leqn{parameters}, we obtain
\beqa
A(\mu) &=& \frac{4\mu^2}{m_Z^2}\,\left(1+\frac{m_A^2+m_Z^2}{m_A^2}
\tan^2 2\beta \right), \CR 
A(b) &=& \left( 1+\frac{m_A^2}{m_Z^2}\right)\tan^2 2\beta, \CR
A(m_u^2) &=& \left| \frac{1}{2}\cos2\beta +\frac{m_A^2}{m_Z^2}\cos^2\beta
-\frac{\mu^2}{m_Z^2}\right|\times\left(1-\frac{1}{\cos2\beta}+
\frac{m_A^2+m_Z^2}{m_A^2} \tan^2 2\beta \right), \CR 
A(m_d^2) &=& \left| -\frac{1}{2}\cos2\beta +\frac{m_A^2}{m_Z^2}\sin^2\beta
-\frac{\mu^2}{m_Z^2}\right|\times\left|1+\frac{1}{\cos2\beta}+
\frac{m_A^2+m_Z^2}{m_A^2} \tan^2 2\beta \right|, \CR 
\eeqa{ders}
where we assumed $\tan\beta>1$. The overall fine-tuning $\Delta$ is defined
by adding the four $A$'s in quadruture; values of $\Delta$ far above one
indicate fine-tuning. For concreteness, we will require $\Delta\leq 100$, 
corresponding to fine tuning of 1\% or better. This requirement maps out
the golden region in the space of $(\tan\beta, \mu, M_A)$, as illustrated in
figure~\ref{fig:FT}. (We do not plot $\mu<100$ GeV, since this region is
ruled out by LEP2 chargino searches.) The shape of this region is easily
understood. In the limit of large $\tan\beta$, the parameters $A(m_u^2)$
and $A(m_d^2)$ are small, and $A(\mu)$ and $A(b)$ (considered separately)
lead to constraints
\beq
\frac{\mu}{m_Z} < \frac{\Delta_{\rm max}^{1/2}}{2},~~~
\frac{m_A}{m_Z} < \frac{\Delta_{\rm max}^{1/2}}{2}\,\tan\beta\,,
\eeq{conlarget}
which are clearly reflected in Fig.~\ref{fig:FT}. As $\beta$ approaches 
$\pi/4$, the factors of $1/\cos2\beta$ and $\tan2\beta$, present in all four 
$A$ parameters, become large, and as a result the model is always fine-tuned 
for $\tan\beta\lsim 2$. 

\subsection{Constraints on the Top Sector}

\begin{figure}[tb]
\begin{center}
\includegraphics[width=5cm]{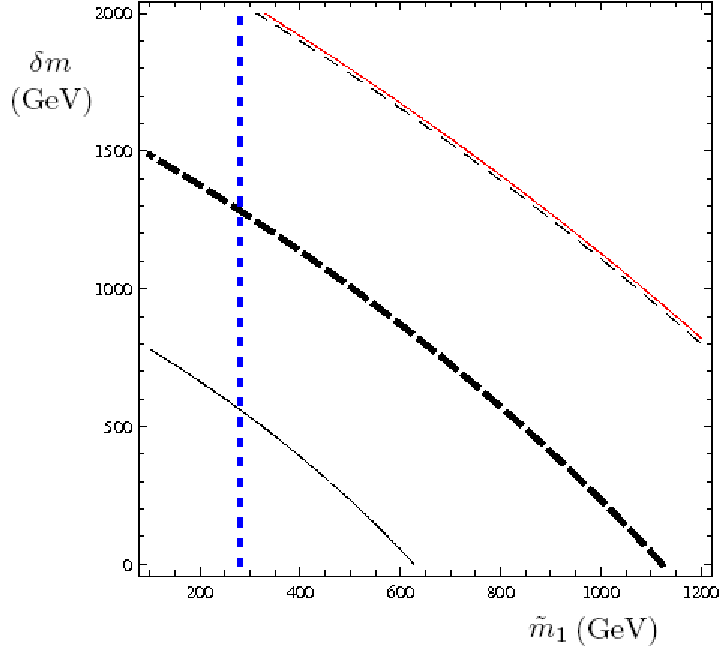}
\includegraphics[width=5cm]{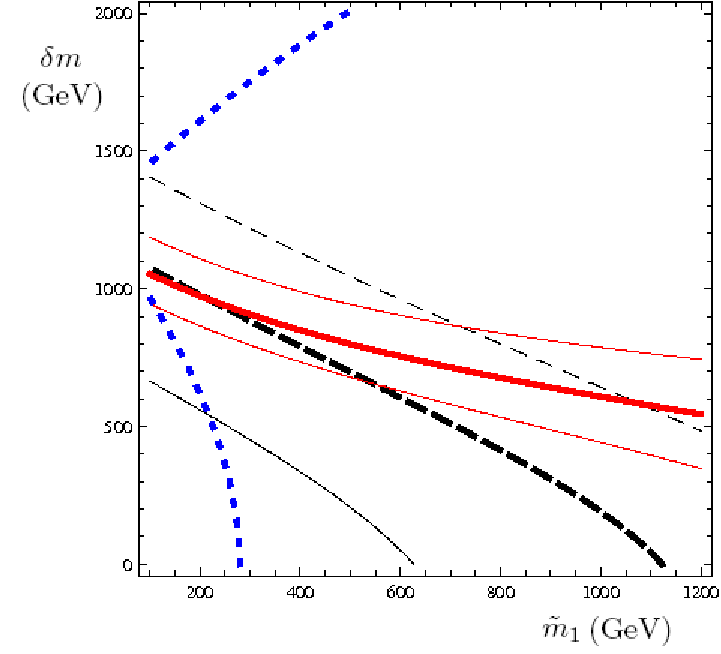}
\includegraphics[width=5cm]{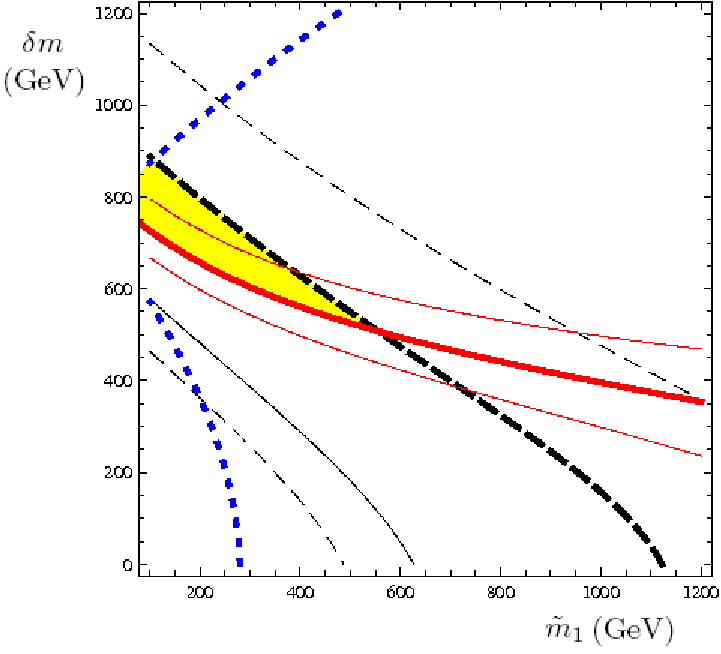}
\includegraphics[width=5cm]{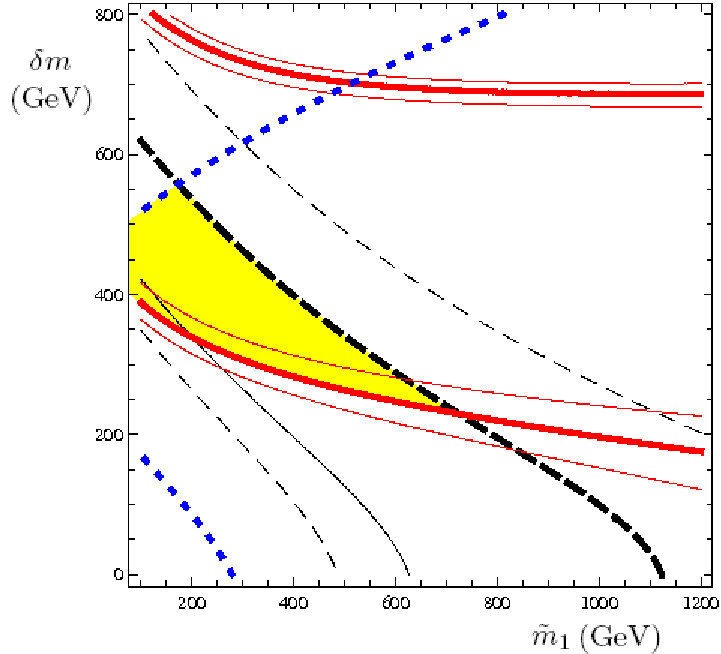}
\includegraphics[width=5cm]{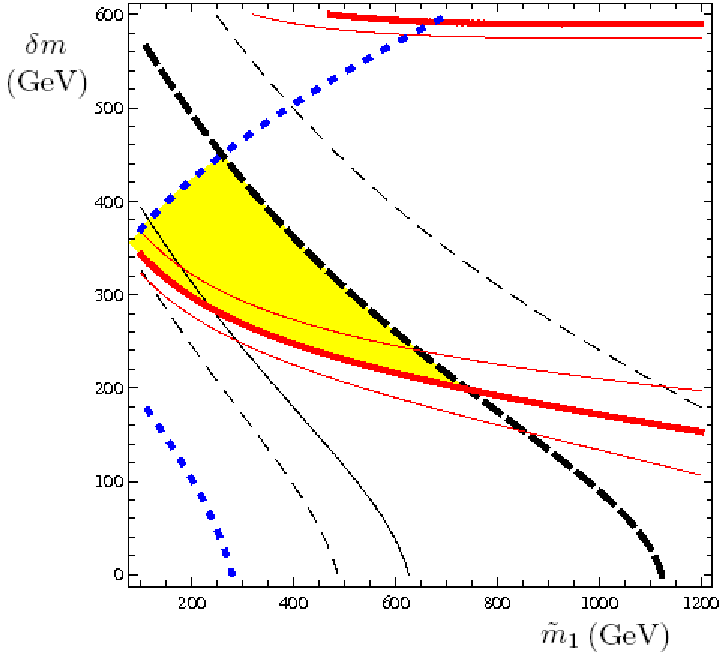}
\includegraphics[width=5cm]{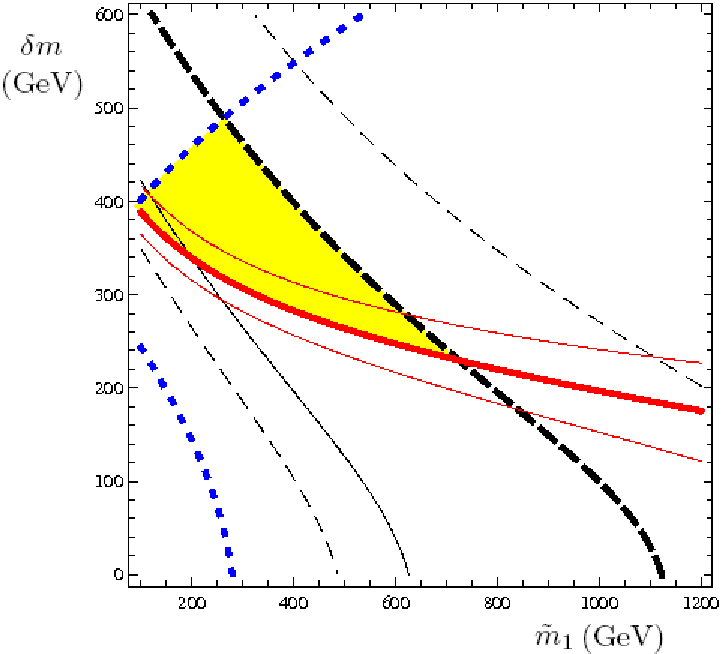}
\vskip2mm
\caption{Fine-tuning (black/dashed contours), Higgs mass bound
(red/colid contours), and $\rho$-parameter (blue/dotted contours)
constraints in the ($\mm$, $\delta m$) plane. The six panels 
correspond to (starting from the upper-left corner, clockwise): $\theta_t=0, 
\pi/25, \pi/15, \pi/6, \pi/4, \pi/3$. In all panels $\tan\beta=10$. The
yellow/shaded intersection of the regions allowed by the three constraints
is the MSSM ``golden'' region.}
\label{fig:FTloop}
\end{center}
\end{figure}

Naturalness also constrains the size of the quantum corrections to the 
parameters in Eq.~\leqn{zmass}. The largest correction in the MSSM is the
one-loop contribution to the $m_u^2$ parameter from top and stop loops:
\beqa
\delta m_{H_u}^2 &\approx & \frac{3 y_t^2}{16\pi^2}\left( 
\tilde{m}^2_{Q_3}+ \tilde{m}^2_{t^c} + 
A_t^2\right) \,\log 
\frac {2\Lambda^2}{\tilde{m}^2_{Q_3}+ \tilde{m}^2_{t^c}} \CR
&\approx & \frac{3}{16\pi^2}\left( y_t^2\left(\mm^2+ \mp^2-2m_t^2\right) + 
\frac{(\mp^2-\mm^2)^2}{4v^2\sin^2\beta}\sin^22\t \right)\,\log 
\frac{2\Lambda^2}{\mm^2+ \mp^2}\,,
\eeqa{toploop}
where $m_t$ is the top mass, $\Lambda$ is the scale at which 
the logarithmic divergence is cut off, and finite (matching) corrections 
have been ignored. In the second line, we re-expressed the correction in terms of the physical (tree-level) stop masses, assuming $A_t\gg \mu/\tan\beta$ (as will always be the case in this study). The correction induced by this effect in the $Z$ mass is 
\beq
\delta_t m_Z^2 \approx -\delta m_{H_u}^2\left( 1-\frac{1}{\cos 2 \beta} 
\right),
\eeq{tloopZ}
where we ignored the renormalization of the angle $\beta$ by top/stop 
loops: the contribution of this effect scales as $1/\tan^2\beta$ and is 
subdominant for $\tan\beta\gsim2$. To measure the
fine-tuning between the bare (tree-level) and one-loop contributions, we
introduce
\beq
\Delta_t \,=\, \left|\frac{\delta_t m_Z^2}{m_Z^2}\right|. 
\eeq{deltat}
Choosing the maximum allowed value of $\Delta_t$ selects a region in the
stop sector parameter space, $(\mm, \mp, \theta_t)$, whose shape is 
approximately independent of the other parameters.\footnote{Note that we 
choose not to combine the tree-level and quantum fine-tuning measures into a 
single tuning parameter; doing so would make the analysis less transparent 
without producing additional physical insights.} This constraint is
shown by the black (dashed) lines in Figs.~\ref{fig:FTloop}, where we plot 
5\%, 3\%, 1\% and 0.5\%
tuning contours (corresponding to $\Delta_t=20, 33.3, 100$, and 200,
respectively) in the stop mass plane for several values of $\theta_t$
and $\tan\beta=10$. Note that the particular values of $\Delta_t$ depend
on the scale $\Lambda$; we choose it to be 100 TeV in this
figure. However, the shape of the contours and the obvious trend for 
tuning to increase with the two stop masses is independent of $\Lambda$.

The second constraint that determines the shape of the golden region is the 
LEP2 lower bound on the Higgs mass~\cite{LEPHiggs}. For generic MSSM parameter
values, the limit on the lightest CP-even Higgs is very close to that for
the SM Higgs:
\beq
m(h^0)\gsim 114~{\rm GeV}.
\eeq{HMlimit}
It is possible for a lighter Higgs (down to about 90 GeV) to be consistent 
with the negative results of the LEP2 searches; however, this requires 
precise coincidence between $m(h^0)$ and $m_A$, which should be regarded as
additional source of fine-tuning. Thus, we will use the LEP2 bound for the
SM Higgs~\cite{PDG}, 114.4 GeV, as the lower 
bound on $m(h^0)$ in this analysis. At tree level, the MSSM predicts
$m(h^0) \leq m_Z|\cos2\beta|$, and large loop corrections are required to 
satisfy this bound. Extensive calculations of these corrections have been
performed in the literature (for a recent summary of the status of these
calculations, see Ref.~\cite{MHfull}). Complete one-loop corrections within
the MSSM are known. The dominant one-loop contribution is from top and stop
loops; for $\tan\beta\gsim 35$, the sbottom loop contribution is also
important. The two-loop corrections to these contributions from strong and
Yukawa interactions are also known. Numerical packages incorporating these
results are available~\cite{FeynHiggs,suspect}. For our purposes here, 
however, it is
convenient to use a simple analytic approximation, due to 
Carena~{\it et.~al.}~\cite{MHanal}, which includes the one-loop and 
leading-log two-loop contributions from top and stop loops: 
\beqa
m^2(h^0)&=& m_Z^2 \cos^2 2\beta\,\left( 1-\frac{3}{8\pi^2}
\frac{m_t^2}{v^2}\,t\right)\CR
&+& \frac{3}{4\pi^2}\frac{m_t^4}{v^2}\left[\frac{1}{2}X_t + t + 
\frac{1}{16\pi^2}\left(\frac{3}{2}\frac{m_t^2}{v^2}-32\pi\alpha_3\right)
\left(X_t t + t^2\right)\right]\,,
\eeqa{mh}
where $\alpha_3$ is the strong coupling constant evaluated at the pole
top quark mass $M_t$; $m_t=M_t/(1+\frac{4}{3\pi}\alpha_3)$ is the on-shell 
top mass; and
\beqa
X_t &=& \frac{2 (A_t-\mu\cot\beta)^2}{\MS^2}\,\left(1-
\frac{(A_t-\mu\cot\beta)^2}{12\MS^2}\right)\,,\CR
t &=& \log \frac{\MS^2}{M_t^2}.
\eeqa{Xt}
The scale $\MS^2$ is defined as the arithmetical average of the diagonal 
elements of the stop mass matrix. The expression~\leqn{mh} is valid when 
the masses of all superparticles, as well as the CP-odd Higgs mass $m_A$, 
are of order $\MS$. Additional threshold corrections may be required, for
example, if $m_A<\MS$; for simplicity, we will ignore such corrections here.
Eq.~\leqn{mh} agrees with the state-of-the-art calculations to within a
few GeV for typical MSSM parameters~\cite{MHfull}; while such accuracy is 
clearly inadequate for precision studies, it is sufficient for the 
present analysis.\footnote{We also verified that the Higgs mass at the 
benchmark point used for the collider phenomenology analysis in this paper 
satisfies the LEP2 bound with a more precise numerical calculation using 
{\tt SuSpect}; see Section 3.}

The contours in the stop mass plane corresponding to the LEP2 Higgs mass 
bound are superimposed on the fine-tuning contours in Figs.~\ref{fig:FTloop}.
The positions of these contours depend strongly on the top quark mass. We used
$M_t=171.4\pm 2.1$ GeV~\cite{mtop}, and plotted the constraint 
corresponding to the central value (thick red/solid lines),
as well as the boundaries of the 95\% c.l. band (thinner red/solid lines). 
The contours are 
approximately independent of $\tan\beta$ for $3\lsim\tan\beta\lsim 35$; 
the golden region shrinks rapidly outside of this range of $\tan\beta$.
We use $\tan\beta=10$ in the plots. 
The overlap between the regions of acceptably low fine-tuning (for
definiteness, we choose $\Delta_t=100$) and experimentally allowed Higgs mass 
defines the golden region, shaded in yellow in Figs.~\ref{fig:FTloop}.

\subsection{Collider Bounds, Precision Electroweak Constraints and Rare 
Decays}

Apart from the Higgs mass bound, several other observables constrain
the shape of the golden region. 

First, direct collider bounds play a role in determining the boundary at 
low $\mu$ and $\mm$: LEP2 searches for direct production of charginos and 
stops constrain both $\mu$ and $\mm$ to be above $\approx 100$ GeV, and are 
to a large extent independent of the rest of the MSSM parameters. (At large 
$\tan\beta$, it can be easily shown that  $m(\chi^\pm_1)< |\mu|$ for any 
$M_2$.) The Tevatron stop searches yield a similar (though more
model-dependent) bound on $\mm$. A sbottom 
search in the $b\chi_1^0$ channel (which is relevant because the 
$\tilde{b}_L$ mass is given by $m_{Q^3}$, and can be expressed in terms 
of $\mm$ and $\mp$) places a lower bound $m(\tilde{b}_L)\geq 200$ GeV.
However, this bound will not be used in our analysis since it is highly 
sensitive to the neutralino mass and can be easily evaded if 
$m(\chi_1^0)>80$ GeV. 

Second, in the presence of a large $A_t$ term, stop and sbottom loops 
may induce a significant correction to the $\rho$ parameter. This correction 
is known at the two-loop level~\cite{rho}; for our purposes, it suffices to
use the one-loop result:
\beq
\Delta\rho \,=\, \frac{3G_F}{8\sqrt{2}\pi^2}\left( -\sin\theta_t^2
\cos\theta_t^2 F_0(\mm^2, \mp^2) + \cos^2\theta_t F_0(\mm^2, m_{\tilde{b}_L}^2)
+ \sin^2\theta_t F_0(\mp^2, m_{\tilde{b}_L}^2)\right)\,,
\eeq{rho}
where
\beq
F_0(a,b) \,=\, a + b -\frac{2ab}{a-b}\log \frac{a}{b}.
\eeq{Fdef}
Expressing $m_{\tilde{b}_L}$ in terms of $\mm$, $\mp$ and $\theta_t$, 
and using the PDG value $\rho = 1.0002  \begin{array}{l} \scriptstyle +0.0004 
\\[-1.5ex] \scriptstyle -0.0007 \end{array}$~\cite{PDG}, we obtain the 
95\% c.l. contours in the stop mass plane shown by the blue/dotted lines in 
Figs.~\ref{fig:FTloop}. This constraint eliminates a part of the parameter 
space with very low $\mm$ and large $\delta m$. 

Finally, several low-energy measurements play a role in constraining the 
MSSM parameter space; among these, the $b\to s\gamma$ decay rate~\cite{bsexp} 
and the anomalous magnetic moment of the muon, $g_\mu-2$~\cite{g2mu}, provide 
the most stringent constraints. The supersymmetric contribution to $g_\mu-2$ 
depends sensitively on the slepton and weak gaugino mass scales, and only 
weakly on the parameters defining the golden region. On the other hand,  
since the golden spectrum contains light stops and higgsinos, we can
expect a large contribution to the $b\to s\gamma$ rate from the $\tilde{t}-
\tilde{H}$ loop. It is well known, however, that this can be cancelled
by the contribution of the top-charged Higgs loop. A simplified analysis
of this constraint based on the one-loop analytic formulas presented in 
Ref.~\cite{bsgamma} shows that for {\it any} values of the stop masses 
inside the golden region in Figs.~\ref{fig:FTloop}, and for any value of
$\mu$ between 100 and 500 GeV, one can find values of $m_A$ in the 
100-1000 GeV range for which this cancellation ensures consistency with 
experiment. (Recall that $m^2(H^\pm)=m_A^2+m_W^2$.)
For low $\mm$ and $\mu$, however, the cancellation only occurs in a narrow
band of $m_A$, which can be thought of as an additional source of fine
tuning. A detailed analysis of this issue is outside the scope of this 
paper.  

\section{A Benchmark Point for Collider Studies}

The analysis of Section 2 defined the golden region in the six-dimensional
parameter space~\leqn{parameters}; its shape is approximately independent of
the other MSSM parameters. This region has the following interesting
qualitative features:
\begin{itemize}

\item Both stops typically have masses below $1$ TeV; 

\item A substantial mass splitting between the two stop quarks is required:
typically, \\ $\delta m\gsim 200$ GeV; 

\item The stop mixing angle must be non-zero: there is no intersection
between the naturalness and Higgs mass constraints for $\theta_t=0, \pi/2$.
\if
, but does not need to be large: 
already for $\theta_t=\pi/15$, there is a sizeable allowed region.
Regions with low $\theta_t$ are characterized by lighter $\tilde{t}_1$
and larger $\delta m$ compared to points with generic $\theta_t$.
\fi
\end{itemize}

\begin{table}[t!]
\begin{center}
\begin{tabular}{|c|c|c|c|c|c|c|c|c|c|c|c|c|} \hline
$m_{Q^3}$ & $m_{u^3}$ & $m_{d^3}$ & $A_t$ & $\mu$ & $m_A$ & $\tan\beta$ 
& $M_1$ & $M_2$ & $M_3$ & $m_{\tilde{q}}$ & $m_{\tilde{\ell}}$ \\ \hline
548.7 & 547.3 & 1000 & 1019 & 250 & 200 & 10 & 1000 & 1000 & 1000 & 
1000 & 1000 \\ \hline
\end{tabular}
\caption{The benchmark point: MSSM input parameters, defined at the weak scale.
(All dimensionful parameters are in GeV.)}
\label{tab:bpdef}
\end{center}
\end{table}

The first feature implies that both $\tl$ and $\th$ will be produced with
sizeable cross sections at the LHC, so that the stop sector can be studied
directly experimentally. The second feature implies that the decay mode
$\th\to \tl Z$ is kinematically allowed. The vertex responsible for this
decay is given by
\beq
\frac{1}{2}
\sin 2\theta_t\,\frac{g}{c_w}\,\left(\frac{1}{2}-\frac{4}{3}s_w^2\right),
\eeq{vertex}
where $c_w$ and $s_w$ are the cosine and sine of the SM Weinberg angle.
The last of the three points above then guarantees that the vertex is 
non-zero, and the decay $\th\to \tl Z$ indeed occurs.

\begin{table}[t!]
\begin{center}
\begin{tabular}{|c|c|c|c|c|c|c|c|c|c|} \hline
$\mm$ & $\mp$ & $m_{\tilde{b}_L}$ \rule[-1.1ex]{0ex}{2ex} & $m(\chi_1^0)$ &
$m(\chi_2^0)$ & $m(\chi_1^\pm)$ & $m_{h^0}$ & $m_{H^0}$ & $m_A$ &
$m_{H^\pm}$ \\ \hline
400 & 700 & 552 & 243 & 253 & 247 & 128.6 & 201 & 200 & 250 \\ 
\hline
\end{tabular}
\caption{The benchmark point: physical spectrum. All masses are in GeV.
The masses of the superparticles not listed here are close to 1 TeV.}
\label{tab:bpspec}
\end{center}
\end{table}

The branching ratio of the $\th\to \tl Z$ mode depends on which competing 
$\th$ decay channels are available. The possible two-body channels 
are
\beq
t \tilde{g},~~~t \tilde{\chi^0},~~~b \tilde{\chi^+}\,,
~~~\tilde{b} W^+,~~~\tilde{b} H^+,~~~\tilde{t}_1 h^0,~~~
\tilde{t}_1 H^0,~~~\tilde{t}_1 A^0\,,
\eeq{chan}
where $\tilde{\chi^0}$ and $\tilde{\chi^+}$ denote all the neutralinos and
charginos that are kinematically accessible, and flavor-changing couplings 
are assumed to be negligible.  

We would like to evaluate the prospects for observing the $\th\to \tl Z$ decay
mode at the LHC. For concreteness, we choose a benchmark point (BP)
within the golden region, and perform a detailed analysis of the signal at 
this point (see Section 4). 
The BP is defined in terms of the weak-scale MSSM parameters.
We assume that all soft parameters are flavor-diagonal.
Further, we assume a common soft mass for the first and second
generation squarks, $m_{\tilde{q}}=m_{Q^{1,2}}=m_{u^{1,2}}=m_{d^{1,2}}$, 
and for all sleptons, $m_{\tilde{\ell}}=m_{L^{1,2,3}}=m_{e^{1,2,3}}=
m_{\nu^{1,2,3}}$. All $A$ terms have been set to zero, with the 
exception of $A_t$. The parameters defining the BP are listed in 
Table~\ref{tab:bpdef}. The top and Higgs sector parameters 
are chosen so that the BP is comfortably inside the golden 
region, well away from the boundaries, and is representative of this region. 
In particular, the lightest Higgs mass at the BP is well
above the LEP bound. (The physical spectrum of the model at the BP
was computed using the {\tt SuSpect} software package~\cite{suspect} and is 
listed in Table~\ref{tab:bpspec}.) Gaugino, slepton, and first and second
generation squark masses are set at 1 TeV. Varying these parameters does
not have a significant effect on the stop production rate and decay 
patterns, and thus the conclusions of the analysis in Section~4 are largely 
independent of these choices. Using {{\tt SuSpect}, we checked that the
$b\to s\gamma$ branching ratio, the $\rho$ parameter and the supersymmetric 
contribution to the muon anomalous magnetic moment at the BP are consistent 
with the current experimental constraints.

\begin{table}[t]
\begin{center}
\begin{tabular}{|c|c|c|c|c|c|c|c|} \hline
$\tl Z$ & $\chi_1^0 t$ & $\chi_2^0 t$ & $\chi_1^+ b$ &
$\tilde{b} W^+$ \rule{0ex}{2.5ex} & $\tl A$ & $\tl h^0$ & $\tl H^0$\\ \hline
31 & 19 & 13 & 18 & 15 & 3 & 3$\times 10^{-3}$& 3$\times 10^{-4}$  \\ 
\hline
\end{tabular}
\caption{The benchmark point: branching ratios of $\th$ decay modes, in \%.}
\label{tab:bran}
\end{center}
\end{table}

The $\th$ decay branching ratios at the BP were evaluated using the 
{\tt SDECAY} package~\cite{sdecay}, and are listed in Table~\ref{tab:bran}. 
The $\th\to\tl Z$ mode has a substantial branching ratio, about 31\%. 
Note that of the possible $\th$ decay modes listed in Eq.~\leqn{chan},
only the $t\tilde{g}$ channel is kinematically forbidden at the BP.
If the gluino mass were lowered to allow this decay, the branching ratio of
the $\th\to\tl Z$ mode would be suppressed. However, this effect is not
dramatic: we checked that if $M_3$ is varied between 300 and 1000 GeV, 
keeping all other MSSM parameters fixed at their values listed in 
Table~\ref{tab:bpdef}, we still obtain Br$(\th\to\tl Z)\gsim 17$\%.
This is an example of the robustness of the stop decay pattern with 
respect to the variations of the non-stop sector MSSM parameters, mentioned
above.

\section{Observability of the $Z + 2 j_b + \met + X$ Signature at the LHC}
\label{obs}

\begin{figure}[tb]
\begin{center}
\includegraphics[width=9cm]{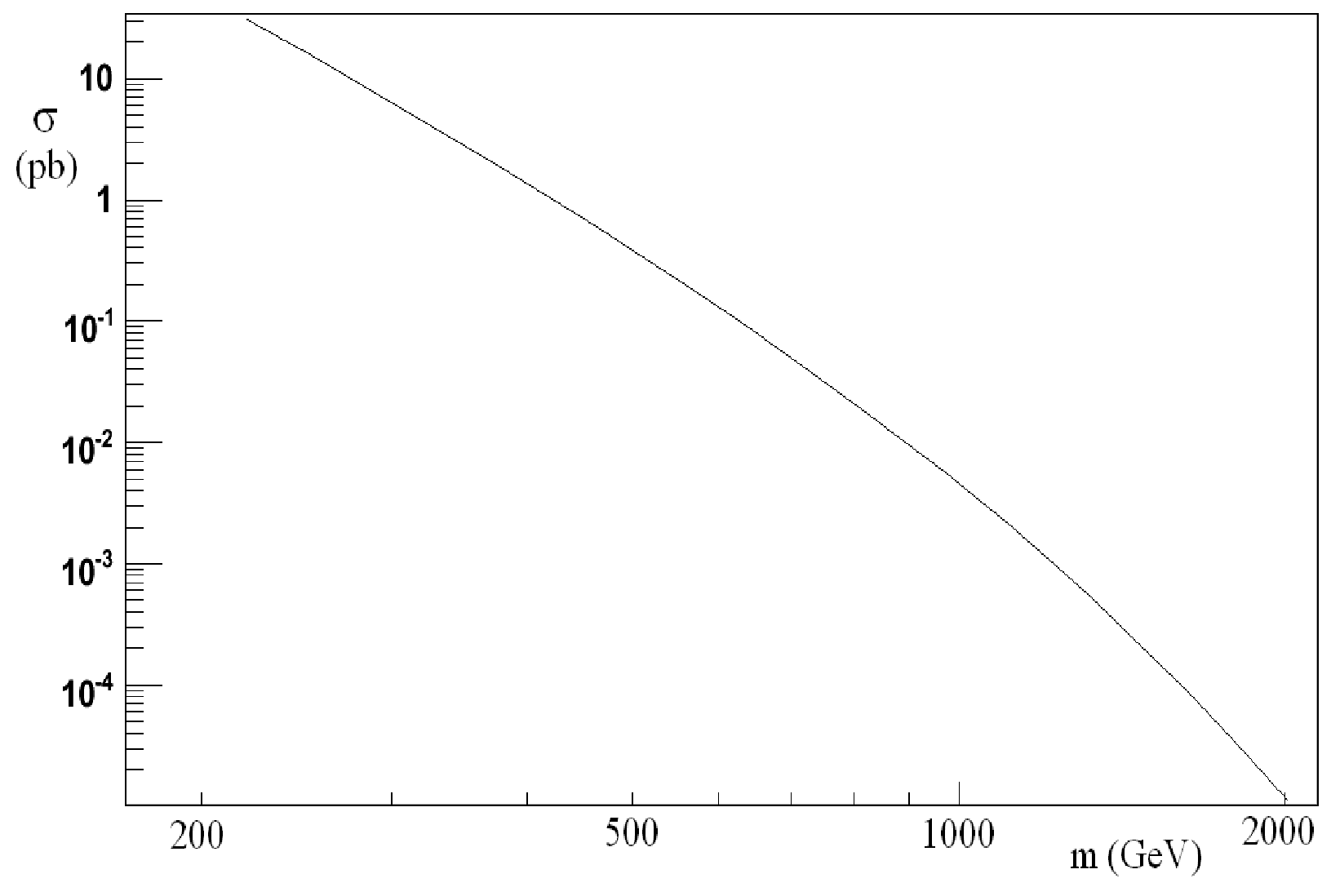}
\vskip2mm
\caption{Cross section of the process $pp\to \tilde{t}\tilde{t}^*$ at the LHC, 
$\sqrt{s}=14$ TeV, at tree level. Factorization and renormalization scales 
were set to $\mu=M_{\tilde{t}}$, and the CTEQ6L1 parton distribution function 
set~\cite{cteq} was used.}
\label{fig:xsec}
\end{center}
\end{figure}

Stop pair production cross section at the LHC, computed using the
{\tt MadGraph/MadEvent v4} software package~\cite{MG}, is shown in 
Fig.~\ref{fig:xsec}. At the
benchmark point, we find $\sigma(pp\to \th \th^*)= 0.05$ pb, corresponding to 
about 500 $\th$ pairs per year at the initial design luminosity of 
10 fb$^{-1}$/year.
The produced $\th$ decays promptly, with branching ratios listed in 
Table~\ref{tab:bran}; in about 52\% of the events, either one or both 
of the produced stops decays in the $\tl Z$ mode. This decay is followed by 
a cascade 
\beqa
\tl \to \chi_1^+ b,~~~\chi_1^+ &\to& u \bar{d} \chi_1^0~/~c \bar{s} \chi_1^0, 
\CR
                             &\to& \ell^+ \nu \chi_1^0, 
\eeqa{cascade}
where the jets and leptons produced in the $\chi_1^+$ decays are very soft 
due to a small chargino-neutralino mass splitting. The details of this 
cascade are particular to the chosen BP, and are quite model-dependent.
There are, however, two model-independent features true for all $\th$ and 
$\tl$ decays: the cascade always contains a $b$ jet 
(produced either directly or via top decay), and it always ends
with the lightest supersymmetric particle (LSP), the neutralino $\chi_1^0$, 
giving a missing transverse energy signature. In order to
make the analysis as model-independent as possible, we
focus on an inclusive signature, 
\beq
Z(\ell^+,\ell^-) + 2 j_b + \met + X,
\eeq{sign}
where $Z(\ell^+,\ell^-)$ denotes a lepton pair ($\ell=e$ or $\mu$) with
the invariant mass at the $Z$ peak. 
The presence of energetic leptons ensures that essentially all such events 
will be triggered on; we will assume a triggering probability of 1 for this 
analysis. Note that events with hadronic $Z$ decays 
may in principle be triggered on due to large $\met$; however, this
sample would suffer from a severe background of purely QCD events with 
apparent $\met$ due to jet energy mismeasurement, and it will not be
used in our study. Note also that the requirement that both jets be 
$b$-tagged can be relaxed, as will be discussed below. While a cleaner 
sample is obtained if two $b$-tags are required, this sample is smaller
due to the less-than-perfect tagging efficiency, which may be relevant 
since the signal rates are not large.

To assess the observability of the signature~\leqn{sign}, 
we have simulated a statistically significant event sample for the signal
and several SM background channels\footnote{
At the chosen benchmark point, the events containing $\th\to\tl Z$ are the only
non-SM source of the signature~\leqn{sign}, so there are no ``signal 
backgrounds''. Possible alternative interpretations of this signature in the
general MSSM context are discussed below in Section~\ref{alt}.}
using the {\tt MadGraph/} {\tt MadEvent v4} software package~\cite{MG}. This 
tool
package allows us to generate both SM and MSSM processes, so that the signal
and backgrounds can be treated uniformly. The parton level events 
generated by
{\tt MadEvent} were recorded in the format consistent with the Les Houches
accord~\cite{LHA,SLHA}. These events were then passed on to the {\tt Pythia} 
package~\cite{Pythia}, which 
was used to simulate showering and hadronization, as well as the decays of 
unstable particles. Finally, the {\tt Pythia} output was processed by the
PGS 3.9 package~\cite{PGS}, which provides a simple and realistic simulation 
of the 
response of a ``typical'' particle detector. (A more detailed analysis of
the detector effects using complete ATLAS and CMS detector simulation 
packages would clearly be interesting, but is outside the scope of this
study.) The final output was analyzed with ROOT, using 
only detector level information for event reconstruction. 

The following SM backgrounds have been considered in detail:

\begin{itemize}

\item $jjZZ$, which can produce the signature~\leqn{sign}
if one $Z$ decays invisibly and the other one is reconstructed
in $\ell^+\ell^-$;

\item $t\bar{t}Z$, with $Z\to \ell^+\ell^-$ and one or both tops decaying 
leptonically (with $\met$ due to neutrinos), or both tops decaying 
hadronically (with $\met$ due to jet energy mismeasurement).

\item $t\bar{t}$, with both tops decaying leptonically and the invariant
mass of the two leptons accidentally close to $m_Z$.

\end{itemize}

The total production cross sections (with $p_{T, {\rm jet}}^{\rm min}=50$ 
GeV for the $jjZZ$ channel) and the
size of the event sample used in our analysis for each channel are listed in 
the first two rows of Table~\ref{tab:cuts}. To identify the events
matching the signature~\leqn{sign}, we impose the following set of 
requirements on the event sample: 

\begin{enumerate}

\item Two opposite-charge same-flavor leptons must be present with 
$\sqrt{s(\ell^+\ell^-)}=M_Z\pm2$ GeV.

\item Two hard jets must be present, with $p_T>125$ GeV for the first jet and
$p_T>50$ GeV for the second jet;

\item At least one of the two highest-$p_T$ jets must be $b$-tagged;

\item The boost factor of the $Z$ boson, $\gamma(Z)=1/\sqrt{1-v_Z^2}$, 
reconstructed from the lepton pair, must be larger than 2.0;

\item A missing $E_T$ cut, $\met>225$ GeV.

\end{enumerate}

\begin{table}[t!]
\begin{center}
\begin{tabular}{|l||r||r|r|r||r|} \hline
& signal: $\tilde{t}_2\tilde{t}_2^*$ & $jjZZ$& $t\bar{t}Z$& $t\bar{t}$& $jjZ$ 
\rule{0ex}{2.2ex} \\ \hline \hline
$\sigma_{\rm prod}$(pb) & 0.051& 0.888& 0.616& 552& 824\\ \hline
total simulated & 9964& 159672& 119395& 3745930& 1397940\\ \hline
1. leptonic $Z$(s) & 1.4& 4.5& 2.6& 0.04& 2.1 \\ \hline
2(a). $p_t(j_1) >125$ GeV & 89& 67& 55& 21& 41\\ \hline
2(b). $p_t(j_2) >50$ GeV & 94& 93& 92& 76& 84\\ \hline
3. $b$-tag & 64 & 8& 44& 57& 5\\ \hline
4. $\gamma(Z)>2.0$ & 89 & 66& 69& 26& 68\\ \hline
5. $\met > 225$ GeV & 48& 2.2& 4.4& 1.7& $<0.9$ (95\% c.l.)\\ 
 & & & & & 0 (ext.) \\ \hline \hline
$N_{\rm exp}$(100 $fb^{-1}$) & 16.4 & 2.8 & 10.8 & 8.8 &  $<177$ (95\% c.l.)\\ 
 & & & & & 0 (ext.) \\ \hline 
\end{tabular} \\[1ex]
\caption{Summary of the analysis of observability of the supersymmetric
golden region signature~\leqn{sign}. First row: Production cross section 
for the signal and background processes at the LHC. Second row: Number of 
Monte Carlo events used in the analysis. Rows 3--8: Cut efficiencies, in\%. 
Last row: The expected number of events for an integrated luminosity of 
100 fb$^{-1}$.}
\label{tab:cuts}
\end{center}
\end{table}

\noindent The efficiencies of these cuts are given in Table~\ref{tab:cuts},
and the $\met$ distribution of the events passing cuts 1--4 is shown
in Figure~\ref{fig:dist}.
While the overall rate of the SM background processes is much higher than the
signal rate, the cuts 1-5 are quite effective in discriminating 
signal from background. Assuming that the search is statistics-limited, 
we estimate that a 3-sigma observation
would require 75 fb$^{-1}$ of data, while a definitive 5-sigma discovery is
possible with 210 fb$^{-1}$. Note that one important contribution to the 
background, from the $t\bar{t}$ channel, can
be effectively measured from data by measuring the event rates with 
dilepton invariant masses away from the $Z$ peak and performing shoulder 
subtraction. This procedure is likely to be statistics-limited. However,
systematic uncertainties in other background contributions could play a role 
in limiting the reach, and should be studied carefully with a more
detailed detector simulation. 

\begin{figure}[tb]
\begin{center}
\includegraphics[width=10cm]{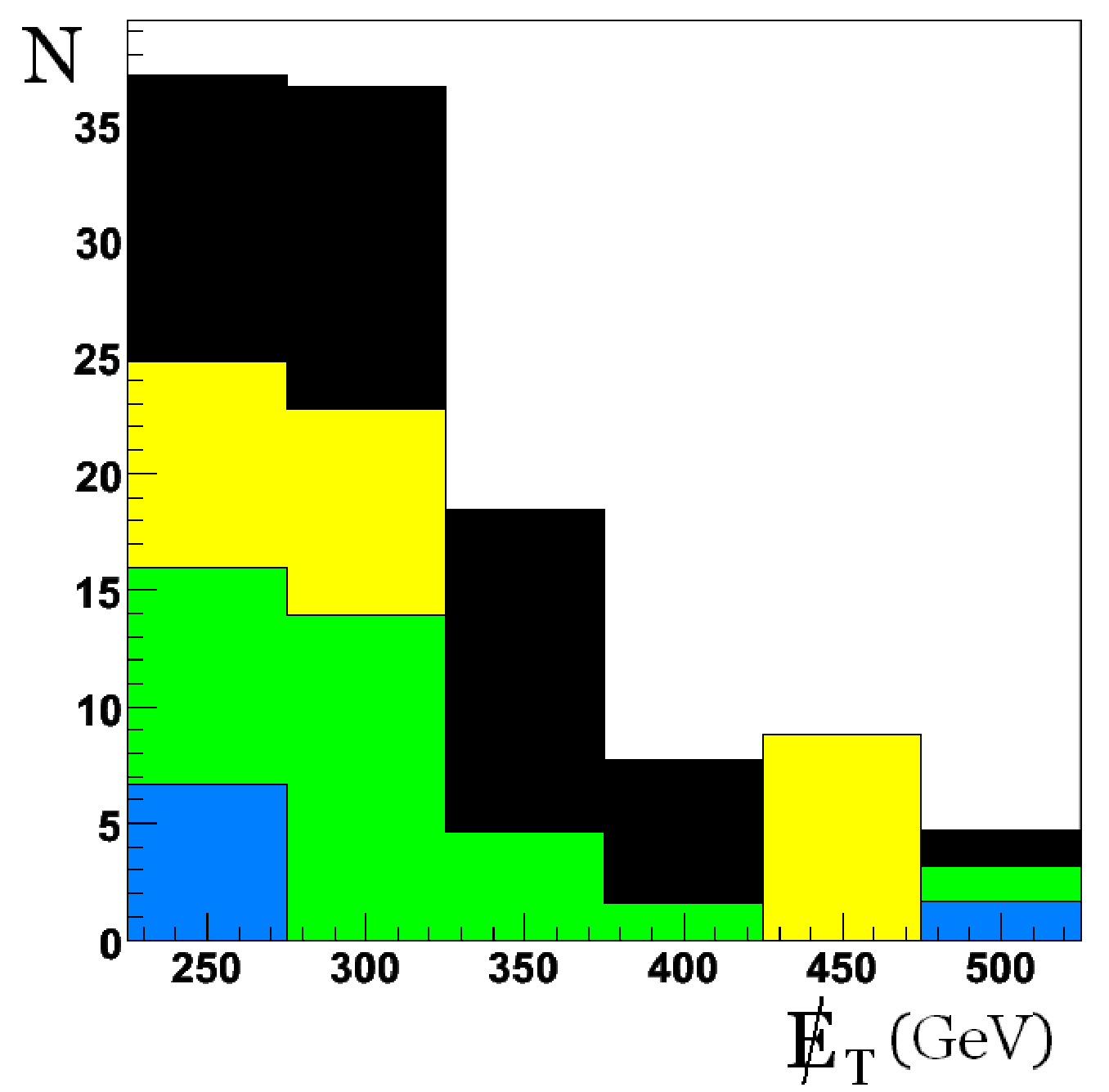}
\vskip2mm
\caption{Missing $E_T$ distribution of the events passing cuts 1--4.
Signal is shown in black; $jjZZ$, $t\bar{t}Z$ and $t\bar{t}$
backgrounds are shown in blue/dark-gray, green/gray, and yellow/light-gray,
respectively.
The normalization corresponds to an integrated luminosity of 300 fb$^{-1}$ 
at the LHC.}
\label{fig:dist}
\end{center}
\end{figure}

We also briefly considered several other irreducible SM backgrounds which
are expected to be less significant than the ones listed in 
Table~\ref{tab:cuts}, but might nevertheless be relevant. The most 
important one of these is $t\bar{t}j$, where $j$ is a hard jet. The cross 
section for this channel is suppressed compared to $t\bar{t}$, but the 
presence of the additional hard jet increases the probability that the events 
will pass the jet $p_T$ cut (cut 2). We find a parton-level cross 
section $\sigma(t\bar{t}j, p_{T}^j>125~{\rm GeV})=65$ pb. Assuming 
conservatively that all these
events pass the cut 2, and that the efficiencies of all other cuts are the 
same as for the $t\bar{t}$ sample, we expect that this background would 
add at most about 50\% to the $t\bar{t}$ rate. As in the $t\bar{t}$ case, 
this contribution can be subtracted using data away from the $Z$ peak in the
lepton invariant mass distribution. Assuming that the statistical error 
dominates
this subtraction, the net effect would be an increase in the integrated 
luminosity required to achieve the same level of significance by at most
about 10\%.\footnote{The $t\bar{t}j$ background may be suppressed very 
effectively by requiring that {\it the hardest} jet be $b$-tagged (as
oppossed to one of the two hardest jets in our main analysis), since the
extra jet is always initiated by a gluon or a light quark. However this
would also reduce the signal and all other backgrounds by about a half due 
to the 
lower probability of tagging a single jet, resulting in lower significance.}
Other backgrounds we considered are three vector boson channels
$ZZZ$, $ZZW$, and $ZWW$; as well as channels with single top production,  
$tZj$ and $\bar{t}Zj$. Combining the parton-level cross sections for these 
channels with the branching ratios of decays producing the 
signature~\leqn{sign} results in event rates that are too small to affect 
the search. 

While the SM processes considered above genuinely produce the 
signature~\leqn{sign}, other SM processes may contribute to the background
due to detector imperfections. We expect that the dominant among these
is the process $jjZ$, with $Z\to \ell^+\ell^-$ and apparent $\met$ 
due to jet energy mismeasurement or other instrumental issues. We
conducted a preliminary investigation of this background by generating 
and analyzing a sample of $1.4\times 10^6$ $jjZ$ events with
$p_{T, {\rm jet}}^{\rm min}=50$ GeV (see the last column 
of Table~\ref{tab:cuts}). None of the events in this sample pass the
cuts 1-5. This allows us to put a 95\% c.l. bound on the combined 
efficiency of
this set of cuts for the $jjZ$ sample of about $2\times 10^{-6}$, 
corresponding to
a background rate about 10 times larger than the signal rate. 
However, we expect that the actual $jjZ$ background rate is well below 
this bound, since all 349 events in our sample that pass the cuts 1--4 
in fact have $\met$ below 50 GeV. We find that the $\met$ distribution
of these 349 events can be fit with an exponential, $N\propto 
e^{-0.10\met}$, where $\met$ is in units of GeV. Assuming that this scaling 
adequately describes the tail of the distribution at large $\met$, we
estimate that the rate of $jjZ$ events passing all 5 cuts is 
completely negligible and that this background
should not present a problem. This conclusion is
of course rather preliminary, and this issue should be revisited 
once the performance of the LHC detectors is understood using real data.
Note that the necessity to understand the shape and normalization of the 
large apparent $\met$ tail from SM processes with large cross sections is 
not unique to the signature discussed here, but is in fact crucial for most 
SUSY searches at the LHC.

\begin{table}[t!]
\begin{center}
\begin{tabular}{|l||r||r|r|r||r|} \hline
& signal: $\tilde{t}_2\tilde{t}_2^*$ & $jjZZ$& $t\bar{t}Z$& $t\bar{t}$& $jjZ$ 
\rule{0ex}{2.2ex} \\ \hline \hline
$\sigma_{\rm prod}$(pb) & 0.051& 0.888& 0.616& 552& 824\\ \hline
total simulated & 9964& 159672& 119395& 3745930& 1397940\\ \hline
1. leptonic $Z$(s) & 1.4& 4.5& 2.6& 0.04& 2.1 \\ \hline
2(a). $p_t(j_1) >125$ GeV & 89& 67& 55& 21& 41\\ \hline
2(b). $p_t(j_2) >50$ GeV & 94& 93& 92& 76& 84\\ \hline
3. 2 $b$-tags & 22 & 0.4& 6& 9& 0.3\\ \hline
4. $\met > 225$ GeV & 56 & $<2$ & $<5$ & 
$<3$ & $<10$ (95\% c.l.)\\ \hline \hline
$N_{\rm exp}$(100 $fb^{-1}$) & 7 & $<2.4$ & $<2.7$ & $<8.8$& $<177$ (95\% c.l.)
\\ 
\hline 
\end{tabular} \\[1ex]
\caption{Same as Table~\ref{tab:cuts}, with an alternative set of 
requirements including 2 $b$-tagged jets.}
\label{tab:cuts2b}
\end{center}
\end{table}

As an alternative, we considered a variation of the analysis where the 
cuts 1, 2, and 5 are unchanged, cut 4 is eliminated, and {\it two}
$b$-tagged jets are required. The cut efficiencies for this analysis 
are summarized in Table~\ref{tab:cuts2b}. Unfortunately, the Monte Carlo
samples used in our analysis are not large enough to reliably estimate 
the efficiencies of this set of cuts applied to the backgrounds, since
only one event out of all background samples passes the cuts. Therefore
we list the 95\% c.l. upper bounds on the efficiencies, and on the
number of background events expected for a 100 fb$^{-1}$ event sample, 
in the table. It is clear that while the second $b$-tag is quite efficient 
in improving the S/B ratio, this search suffers from low statistics,
with only 7 signal events expected in a 100 fb$^{-1}$ data sample.

To summarize, our analysis indicates that, for the MSSM parameters 
at the benchmark point, the signature~\leqn{sign} of the 
split-stop spectrum can be discovered at the LHC. The chosen BP is typical of
the golden region, and this conclusion should generally hold as the 
MSSM parameters are varied away from the BP, scanning this region. There are,
however, several exceptional parts of the parameter space where the
observability of this signature could be substantially degraded. These include:
\begin{itemize}
\item Large $\mp$ region: The $\th$ production cross section drops rapidly 
with its mass, see Fig.~\ref{fig:xsec}, suppressing the signal rates;

\item Small $\theta_t$ region: While non-zero 
$\theta_t$ is required in the golden region, values as small as 
$\theta_t=\pi/15$ are allowed (see Fig.~\ref{fig:FTloop}). The branching 
ratio Br($\th\to Z\tl$) is proportional to $\sin^2 2\theta_t$, see 
Eq.~\leqn{vertex}, and the event rate is suppressed at small $\theta_t$;

\item Small $\tl$-LSP mass difference: The absence of hard jets in this case 
would make the signal/background discrimination more difficult.

\end{itemize}

In these special regions, observing the signature~\leqn{sign} may not be 
feasible at the LHC.
These limitations should be kept in mind when theoretical interpretation of
a search for the signature~\leqn{sign} is given.

\section{Alternative Interpretations of the $Z + 2 j_b + \met + X$ Signature} 
\label{alt}

Unfortunately, observing an excess of events in the channel~\leqn{sign} 
at the LHC does not prove that the decay $\th\to\tl Z$ is occuring. Even
within the MSSM, this is not the only possible interpretation of such an 
excess. The simplest alternative interpretation is stop or sbottom 
production, followed by a cascade decay containing a $b$ quark and
a $Z$ boson from a neutralino or chargino decay: $\chi^0_j\to\chi_i^0 Z$ 
(due to the higgsino components of the neutralinos) or
$\chi^\pm_2 \to \chi^\pm_1 Z$. Can this alternative interpretation be 
ruled out based on data?

One useful input for discriminating between these two interpretations is 
whether a signal is observed in a search identical to the one presented in
Section~\ref{obs}, but requiring that the jets {\it not} be $b$-tagged. If the
signal is due to $\th\to\tl Z$, all signal events contain energetic $b$ 
quarks, and the number of events in this search would be zero if $b$
tagging were perfect. The actual number of expected events under realistic 
conditions can be deduced from the error rate in $b$ tagging, which can be
measured elsewhere. If the signal is due to $\chi^0_j\to\chi_i^0 Z$ or
$\chi^\pm_2 \to \chi^\pm_1 Z$, it does not have to be associated 
preferentially 
with third-generation squark production, and the number of events without $b$ 
tags could be substantially larger than this expectation. This argument
could be used to rule out the $\th\to\tl Z$ interpretation. Unfortunately,
however, it cannot be used to confirm it: the pattern consistent with
the $\th\to\tl Z$ interpretation may also appear if the events are 
actually due to chargino or neutralino decays, provided that the first two 
generations of squarks are substantially heavier than their counterparts of 
the third generation and their production cross section is suppressed.
A direct measurement of the squark masses could break this degeneracy. 
If the first two generations of squarks were found to be light, but 
no signal is seen in $Z+2j +\met$ with non-$b$ jets, the ``split stop''
interpretation of the signal~\leqn{sign} would be preferred.

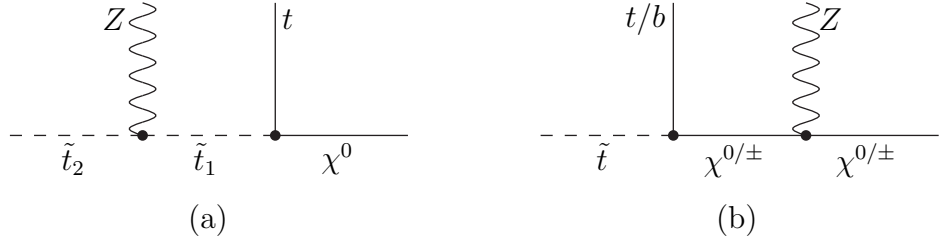
\begin{figure}[t]
\begin{center}
\begin{picture}(360,60)(0,-25)
\Text(25,-2)[tc]{$\th$}
\Text(75,-2)[tc]{$\tl$}
\Text(125,-2)[tc]{$\chi^0$}
\Text(45,45)[cr]{$Z$}
\Text(104,45)[cl]{$t$}
\Text(76,-25)[tc]{(a)}
\DashLine(1,1)(51,1){5}
\Photon(51,51)(51,1){5}{5}
\Vertex(51,1){2}
\DashLine(51,1)(101,1){5}
\Vertex(101,1){2}
\Line(101,1)(101,51)
\Line(101,1)(151,1)
\Text(225,-2)[tc]{$\tilde{t}$}
\Text(275,-2)[tc]{$\chi^{0/\pm}$}
\Text(325,-2)[tc]{$\chi^{0/\pm}$}
\Text(249,45)[cr]{$t/b$}
\Text(307,45)[cl]{$Z$}
\Text(276,-25)[tc]{(b)}
\DashLine(251,1)(201,1){5}
\Line(251,1)(251,51)
\Vertex(251,1){2}
\Line(301,1)(251,1)
\Vertex(301,1){2}
\Photon(301,51)(301,1){5}{5}
\Line(301,1)(351,1)
\end{picture}
\vskip2mm
\caption{Cascade decays in the MSSM leading to the $Z+2j+\met$ signature:
(a) the chain characteristic of the golden region; (b) an 
alternative chain.}
\label{fig:diagrams}
\end{center}
\end{figure}

A more direct way to discriminate between the two interpretations
would be to study the distribution of the events as a function of 
the $Z$-jet invariant mass $s_{jZ}\equiv(p_j+p_Z)^2$. This strategy is the 
same as the recently
proposed method of discriminating between SUSY and alternative theories with
same-spin ``superpartners''~\cite{cascade,WY}, but in this case it is applied
to distinguishing two processes within the MSSM. Consider the Feynman diagrams
corresponding to the two interpretations of the signal, shown in
Figure~\ref{fig:diagrams}. In the case of $\th\to\tl Z$ 
decays, the $Z$ and the jet are separated by a scalar (stop) line, and their 
directions are uncorrelated. In the case of chargino or neutralino decays,
the $Z$ and the jet are separated by a fermion line, and 
spin correlations between their directions are possible. Unfortunately,
in the neutralino case, no such correlations occur, because of the non-chiral
nature of the $\chi^0_i \chi^0_j Z$ coupling~\cite{WY}. In the chargino
case, however, the coupling has the form~\cite{HK}
\beq
-\frac{g}{2c_w}\,\bar{\tilde{\chi}}_i
\gamma^\mu\,\left[C^L_{ij} (1-\gamma_5) \,+\, C^R_{ij} (1+\gamma_5) 
\right] \tilde{\chi}_j\,,
\eeq{xxZ}  
where 
\beqa
C^L_{ij} &=& V_{i1} V_{j1}^* + \frac{1}{2} V_{i2} V^*_{j2}\,, \CR
C^R_{ij} &=& U_{i1}^* U_{j1} + \frac{1}{2} U_{i2}^* U_{j2}\,.
\eeqa{CLCR}
Here, $U$ and $V$ are the rotations of the negatively-charged and 
positively-charged charginos, respectively, required to diagonalize 
the chargino mass matrix. Since in general $U\not= V$, the couplings~\leqn{xxZ}
are generically chiral. The stop-bottom-chargino coupling is also 
generically chiral: it has the form
\beq
\bar{b} (A^L_{ij}P_R + A^R_{ij} P_L) \tilde{\chi}^c_i \tilde{t}_j,
\eeq{tbx}
which can be equivalently rewritten as
\beq
\bar{\tilde{\chi}}_i(A^L_{ij}P_R + A^R_{ij} P_L) b^c \tilde{t}_j. 
\eeq{tbx_alt}
Here
\beqa
A^L_{ij} &=& g V_{i1} R^t_{j1} + \frac{y_t}{s_\beta} V_{i2} R^t_{j2}\,, \CR
A^R_{ij} &=& \frac{y_b}{c_\beta} U_{i2} R^t_{j1} \,,
\eeqa{ALAR}
where $R^t$ is the matrix diagonalizing the stop masses: 
$(\tl,\th)^T = R^t (\tilde{t}_L,\tilde{t}_R)^T$.
Squaring the matrix element ${\cal M}$ for the decay 
$\tilde{t}_j\to b+Z+\chi^+_1$, see Fig.~\ref{fig:diagrams}~(b), and summing 
over the final-state polarizations yields
\beq
\sum_{{\rm pol}}|{\cal M}|^2 \,\propto (|A^L_{j2}|^2-|A^R_{j2}|^2)\,
(|C^L_{12}|^2-|C^R_{12}|^2)\,s_{bZ} + 
{\rm const},
\eeq{correl}
where the constant terms do not depend on $s_{bZ}$, narrow-width 
approximation for $\chi^\pm_2$ has been used, and the $b$ quark mass was 
neglected. The charge-conjugate decay $\tilde{t}^*_j\to \bar{b}+Z+\chi^-_1$
has the same asymmetry. Observing a linear dependence of the event rate on 
$s_{bZ}$ would provide clear evidence against the interpretation of the signal
in terms of the process in Fig.~\ref{fig:diagrams} (a). Of course, in a
real experiment, the asymmetry would be partially washed out by combinatoric
backgrounds, as well as possible non-chiral decay chains containing the 
same final state. A detailed analysis of the observability of the 
correlation in Eq.~\leqn{correl} is beyond the scope of this paper. 

While our analysis so far focused on the decay $\th\to\tl Z$ as a signature
of the MSSM golden region, there are two other, closely related decays that
are also characteristic of this region:
\beq
\th \to \tilde{b}_L W^+,~~~\tilde{b}_L \to \tl W^-.
\eeq{cousin} 
For example, at the benchmark point used for the analysis in Section 4,
these two decays have branching ratios of 15\% and 43\%, respectively.
Stop or sbottom pair-production followed by these decays leads to a 
signature 
\beq
W+2j_b+\met+X.
\eeq{cousin_sign}
This signature is complementary to the $Z+2j_b+\met+X$ signature studied above.
On the one hand, it suffers from higher backgrounds, since the $W$ cannot be
fully reconstructed in purely leptonic channels. On the other hand, its 
interpretation within the MSSM is somewhat cleaner. The leading 
alternative interpretation of the signature~\leqn{cousin_sign} is that
the $W$'s are produced in chargino $\to$ neutralino decays. But the 
chargino-neutralino coupling is chiral, and the directions of the 
$W$ and the associated jet are correlated. If the $W$ is sufficiently boosted, 
this will result in an observable linear dependence of the cross section
on $s_{\ell j}\equiv (p_\ell+p_j)^2$, where $\ell$ is the lepton daughter of 
the $W$~\cite{WY}. 
If, on the other hand, the $W$ is produced in decays of scalars, such as
the processes~\leqn{cousin}, the distribution of events in $s_{\ell j}$
should be flat. 

To summarize, even if the MSSM is assumed to be the underlying model, the 
interpretation of events with
vector bosons associated with jets and missing $E_T$ is not unambiguous. 
Careful comparisons of the rates with and without $b$ jets, as well as 
the distribution of events in vector boson-jet invariant masses, would
be required to remove the ambiguity. This may take considerably more data
than the discovery of an excess over the SM backgrounds in these channels.

If the MSSM is {\it not} assumed from the beginning, the question of 
interpretation becomes even more confusing. For example, in the Littlest 
Higgs model with T-parity~\cite{LHT}, $Z$ bosons can be produced in the
decay $W^3_H\to Z B_H$, due to the mixing between the $SU(2)$ and $U(1)$ 
heavy gauge bosons. A similar decay involving the Kaluza-Klein states of 
the $SU(2)\times U(1)$ gauge bosons can occur in models with universal
extra dimensions (UED)~\cite{UED}. Again, a careful study of spin 
correlations 
would be necessary to disentangle these possibilities. Understanding the
nature of such correlations in various models is an interesting direction for 
future work.  

\section{Conclusions}

In this paper, we discussed an LHC signature of the MSSM characteristic of
the ``golden region'' in the model parameter space. The advantage of this
signature is that it directly probes the features of the stop spectrum that
are dictated by naturalness and the Higgs mass bound. Experimentally, the
signature is not straightforward, but the results of our simulations 
indicate that it should be within reach at the LHC. 

Given the strong theoretical motivation for the signature discussed here,
we encourage experimental collaborations to perform a more detailed 
study of its observability. The analysis of this paper relied on a
set of simple rectangular cuts, and no systematic procedure to optimize the
cuts was employed. It is very likely that a better algorithm for
signal/background discrimination, perhaps using modern data analysis tools 
such as neural networks or decision trees, would significantly enhance the 
reach. On the other hand, it should be noted that we ignored systematic 
uncertainties on the background rates in our reach estimates, and that no 
fully realistic detector simulation was attempted. 

If the first round of the LHC results points towards an MSSM-like theory,
obtaining experimental information about the stop spectrum, and in particular
testing whether the ``golden region MSSM'' hypothesis is correct, will
become an important priority for the LHC experiments. An indirect way to
shed some light on this issue by identifying the stop loop contributions to 
the Higgs production cross section and mass has been recently proposed
by Dermisek and Low~\cite{DL}. This is complementary to the direct
probe explored in this paper. It would be interesting to explore other
experimental consequences of the golden region hypothesis. 

\vskip0.5cm

\noindent{\large \bf Acknowledgments} 

Giacomo Cacciapaglia has collaborated with us 
at the early stages of this project. We gratefully acknowledge his 
contributions to our understanding of the implications of naturalness and 
experimental data on the MSSM parameters. We are grateful to Patrick
Meade and Michael Peskin for useful discussions.  
This research is supported by the NSF grant PHY-0355005.

\end{document}